\documentclass{aastex701}
\usepackage{nameref}

\begin{document}

\title{Observational Evidence Linking Loop Length and Thermal-Nonthermal Peak Timing in Solar Flares}

\author[orcid=0009-0002-6257-3034,sname='Solomon M. Perriyil']{Solomon M. Perriyil}

\affiliation{Center for Radio Astronomy and Astrophysics Mackenzie, Mackenzie Presbyterian University, Sao Paulo, Brazil}
\email[show]{sperriyil31@gmail.com}  

\author[orcid=0009-0008-9486-6417,sname='Soumya S. Sadangaya']{Soumya S. Sadangaya}

\affiliation{Center for Radio Astronomy and Astrophysics Mackenzie, Mackenzie Presbyterian University, Sao Paulo, Brazil}
\email[show]{soumyashree0312@gmail.com} 

\author[orcid=/0000-0002-8979-3582,sname=C. Guillermo Gim\'enez de Castro]{C. Guillermo Gim\'enez de Castro}
\affiliation{Center for Radio Astronomy and Astrophysics Mackenzie, Mackenzie Presbyterian University, Sao Paulo, Brazil}
\affiliation{Institute of Astronomy and Space Physics, UBA/CONICET, Buenos Aires, Argentina}
\email[show]{guigue@craam.mackenzie.br}

\author[orcid=/0000-0002-4819-1884,sname=Paulo J. A. Sim\~oes]{Paulo J. A. Sim\~oes}
\affiliation{Center for Radio Astronomy and Astrophysics Mackenzie, Mackenzie Presbyterian University, Sao Paulo, Brazil}
\affiliation{SUPA School of Physics and Astronomy, University of Glasgow, Glasgow, UK}
\email[show]{paulo@craam.mackenzie.br}

\begin{abstract}
We investigate how the magnetic loop length of solar flares relates to the timing between their thermal and nonthermal emission signatures. Our study analyzes a sample of 96 C, M, and X-class flares observed between 2013 and 2015 with soft, hard X-rays and extreme UV. For each event, we determine the time delay $\Delta t$ between the HXR and the SXR peak, and estimate the flare loop length $L$ from UV footpoints assuming a semicircular geometry. \added{In every case, longer flare loops are consistently associated with larger timing delays.} Across the full sample, we find a strong correlation $R = 0.88$ between $L$ and $\Delta t$. \added{We also quantify how closely each flare follows the Neupert effect using a coefficient $R_{\mathrm{N}}$, defined as the Pearson correlation between the time derivative of the soft X-ray flux and the hard X-ray light curve.} Applying correlation thresholds of $R_{\mathrm{N}} \geq 0.5$ and $R_{\mathrm{N}} \geq 0.8$ yields subsets of 87 and 46 events, respectively. In both cases, the linear relationship between loop length and peak delay remains clearly expressed. For the $R_{\mathrm{N}} \geq 0.5$ subset, the correlation is $R = 0.87$, while the more selective subset with $R_{\mathrm{N}} \geq 0.8$ displays an even stronger correlation of $R = 0.91$. These results show that the overall trend persists across increasingly stringent correlation thresholds. These results provide direct observational confirmation that magnetic loop geometry plays a key role in governing the temporal evolution of energy transport in solar flares.
\end{abstract}

\keywords{Sun: flares --- Sun: corona --- Sun: X-rays, gamma rays --- Sun: magnetic fields --- Sun: chromosphere}


\section{Introduction} 

Solar flares are sudden and powerful energy release events in the solar atmosphere that convert stored magnetic energy into plasma heating, particle acceleration, and radiation across the electromagnetic spectrum \citep{PriestForbes2002}. Magnetic reconnection in the corona \citep{LinForbes2000, PriestForbes2002} accelerates electrons, which travel along magnetic field lines toward the dense chromosphere. Through Coulomb collisions, these energetic electrons deposit their energy into the lower atmosphere, producing hard X-Rays (HXR) nonthermal bremsstrahlung \citep{Brown1971, Emslie1978}. The energy deposited by nonthermal electrons heats the chromosphere, causing the heated material to expand upward along the magnetic loop, a process known as chromospheric evaporation \citep[e.g.][]{Nagai1980,  Fisher1985a, Fisher1985b, Mariska1989}. During this process, dense chromospheric plasma is rapidly heated to temperatures of several million \added{Kelvin} and rises into the corona, filling the newly reconnected magnetic loop with hot, high pressure plasma. As the loop becomes filled and begins to cool, it emits soft X-Rays (SXR). The resulting SXR emission therefore provides a direct observational signature of the plasma’s thermal response to the earlier nonthermal electron energy deposition.

 In hydrodynamic models based on the collisional thick-target scenario, flares occurring in longer magnetic loops develop larger delays between the HXR peak and the subsequent SXR peak, because the conductive heat front and evaporated plasma must travel farther before the coronal loop fills and begins to cool \citep{Fisher1985a, Fisher1985b}. \added{Here, chromospheric heating generates an upward conductive front.} Later simulations quantified how heating, transport, and cooling timescales increase with loop length and depend on plasma conditions, showing that long loops remain hot for longer and therefore delay the rise of SXR emission relative to the impulsive HXR phase \citep{CargillMariskaAntiochos1995}. This naturally explains why nonthermal emission often peaks earlier than the thermal emission and provides the physical basis for interpreting HXR-SXR timing delays in statistical flare studies closely linked to what is commonly known as the Neupert effect \citep[e.g.][]{Neupert1968, Veronig2002a,Veronig2002b}.

\added{The Neupert effect} describes the tendency for the time integral of the HXR or microwave emission which traces accumulated energy deposited by nonthermal electrons to closely match the gradual rise of the SXR emission, which reflects the thermal plasma heating. This relationship suggests a causal connection: the same nonthermal electrons that produce HXR bremsstrahlung as they precipitate into the chromosphere are also responsible for heating the plasma, which subsequently evaporates into the corona and emits in SXR \citep[e.g.][]{DennisZarro1993, Veronig2002a,Veronig2002b, WarmuthMann2016}. However, this behavior is not universal. Many flares show departures from the Neupert effect, which can arise from additional or prolonged energy release, early gradual heating (preheating) before electron acceleration, or conduction-driven heating rather than direct electron beam energy deposition \citep{Veronig2002a,Veronig2002b}. \added{In this scenario, coronal energy conducts downward, causing an early SXR increase.} These deviations underline that the energy transport mechanisms and heating pathways can vary substantially between flares, making the Neupert effect a powerful but not guaranteed diagnostic of electron-beam driven energy release.

Previous statistical studies have investigated the timing relationship between SXR and HXR emissions as a diagnostic of energy transport. \citet{LiGan2006} analyzed 859 flares using GOES and BATSE data and found that the distribution of SXR-HXR delays closely resembled the distribution of flare loop lengths inferred independently by \citet{Aschwanden1998}, suggesting a possible geometric origin for the delay. However, neither GOES nor BATSE provided spatial imaging, and thus loop lengths could not be directly measured. Consequently, this proposed link between loop geometry and emission timing has not yet been conclusively verified using simultaneous temporal and spatial observations.

In this study, we present the first imaging based statistical analysis that directly relates flare loop length to the temporal delay ($\Delta t$) between thermal and nonthermal emission peaks. We analyze 96 solar flares observed between 2013 and 2015, using coordinated data from GOES, RHESSI, and SDO/AIA, comprising 5 X-class, 62 M-class, and 29 C-class events. For each event, the SXR and HXR peak times are determined, $\Delta t$ is calculated, and the corresponding loop length is measured from imaging data. In addition, we examine whether this relationship between loop length and timing delay \added{is associated with} the Neupert effect by comparing results from the full sample with those flares that exhibit a greater Neupert correlation. In the following sections, we first describe the datasets and selection criteria used in this study (2) and outline the methods adopted to determine the peak timings and magnetic loop lengths (3). The statistical results and their relation to the Neupert effect are presented in (4), followed by a discussion of their physical implications in (5). Finally, we summarize our conclusions in (6).

\section{Instrumentation and Flare Selection} \label{Instrumentation and Flare Selection}
SXR fluxes were obtained from the Geostationary Operational Environmental Satellite (GOES), which monitors solar emission in two wavelength bands, 0.5-4 Å and 1-8 Å.
For this analysis, the 1–8 Å channel was used, as it traces the thermal emission from hot coronal plasma. The data were provided by the NOAA National Centers for Environmental Information (NCEI).

HXR observations were taken from the Reuven Ramaty High Energy Solar Spectroscopic Imager (RHESSI) \citep{Lin2002RHESSI}, which measures solar X-rays and gamma rays over an energy range of 3 keV-17 MeV with high spectral and temporal resolution.
We focus on the 25-50 keV energy channel, which primarily represents nonthermal bremsstrahlung from accelerated electrons. These data were also used to produce CLEAN images to locate HXR footpoints and reconstruct the flare geometry.

Ultraviolet (UV) observations were obtained from the Solar Dynamics Observatory (SDO) \citep{Pesnell2012}.
Specifically, the Atmospheric Imaging Assembly (AIA) was used in its 1700 Å channel, which samples lower chromospheric emission and is well suited for identifying flare ribbons and footpoints.
The AIA 1700~\AA\ images have a cadence of 24~s and a diffraction limited spatial resolution of $\approx$1.5\arcsec\  \citep{Lemen2012AIA}.

These images were co-aligned with RHESSI CLEAN maps to compare the spatial distribution of nonthermal and chromospheric sources and to measure footpoint separation.

To ensure accurate timing and reliable imaging, only flares that met the following criteria were selected:

- Events were required to lie within ±60° heliographic longitude of the solar disk center to minimize projection effects in loop length estimation.  

- Each flare was required to show a clear, high signal to noise ratio in both GOES and RHESSI light curves, particularly in the 25-50 keV channel, enabling precise determination of the nonthermal emission peak. For multi-peaked events, the strongest HXR peak was selected for analysis.

- Only flares with sufficient RHESSI count statistics to generate stable CLEAN images specifically in the 25-50 keV band were included. CLEAN reconstructions were used to locate the nonthermal footpoint sources, and all imaging intervals were centered on the RHESSI HXR peak.

- Events were required to have overlapping observations from GOES, RHESSI, and SDO/AIA, ensuring consistent temporal and spatial coverage across all wavelengths.

\begin{figure}[ht!]
    \centering
    \includegraphics[width=0.65\linewidth]{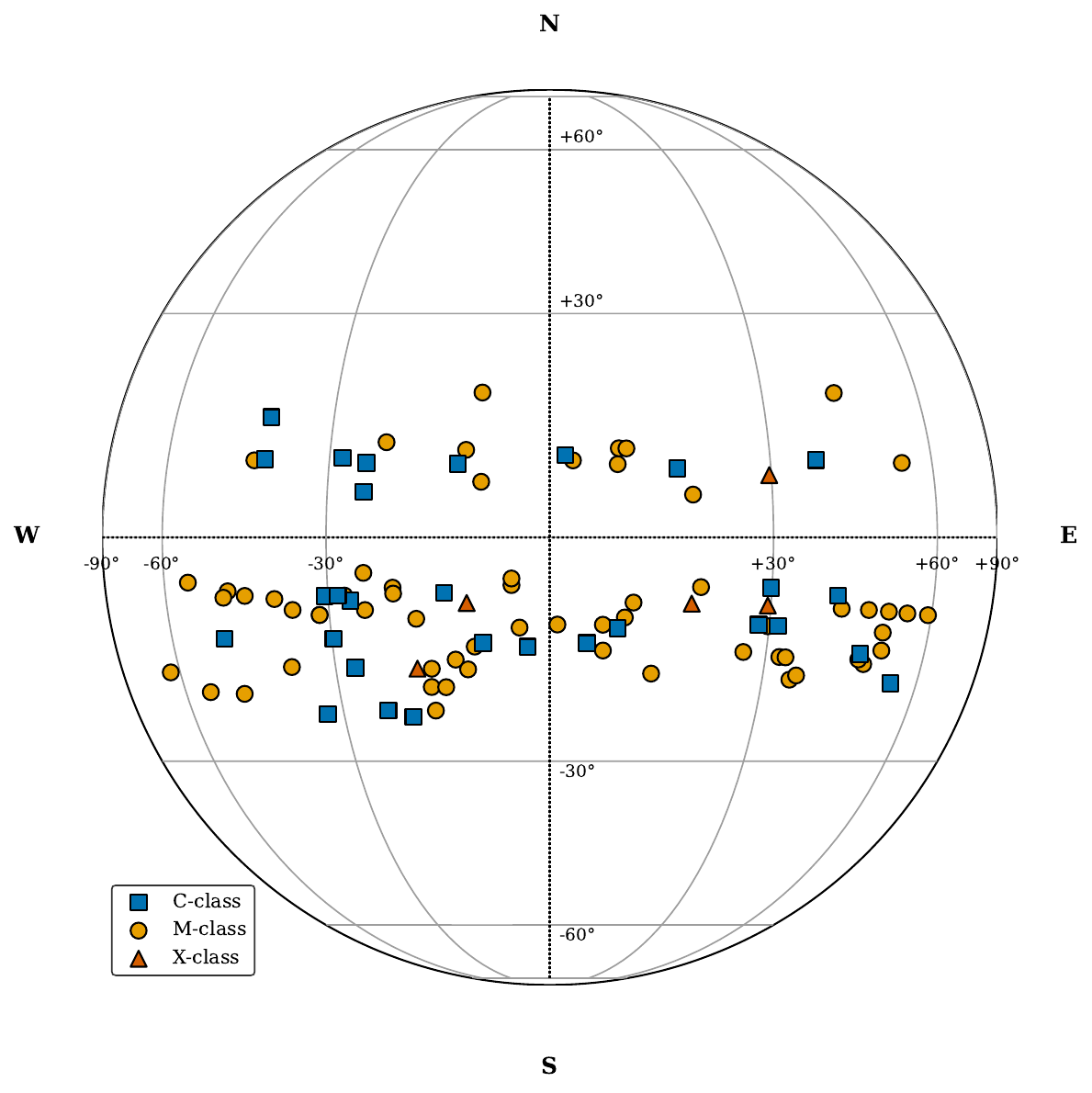}
    \caption{Heliographic positions of the 96 selected flares. 
    All events lie within $\pm60^\circ$ heliographic longitude of the solar-disk center,
    ensuring minimal projection effects in loop-length estimation. 
    square, circle, and triangle markers denote C, M, and X-class flares, respectively.}
    \label{fig:flare_loc}
\end{figure}

After applying these criteria, a total of 96 high quality events remained. This dataset provides a reliable and statistically meaningful foundation for investigating the relationship between the temporal delay ($\Delta t$) between SXR and HXR peaks and the magnetic loop length. The Heliographic positions of the 96 selected flares are shown in Figure~\ref{fig:flare_loc}

\section{Methodology} \label{sec:floats}
\subsection{Determination of SXR and HXR Peak Times}

To study the timing between thermal and nonthermal flare emissions, we used SXR data from the GOES 1-8 Å channel and HXR data from the RHESSI 25-50 keV channel. 
The GOES 1-8 Å light curve traces the thermal emission from hot coronal plasma produced by chromospheric evaporation, while the RHESSI 25-50 keV light curve represents nonthermal bremsstrahlung emission generated by accelerated electrons. 
Together, these two diagnostics allow us to quantify the timing of energy release and plasma response during solar flares.

\begin{figure}[ht!]
    \centering
    \includegraphics[width=0.65\linewidth]{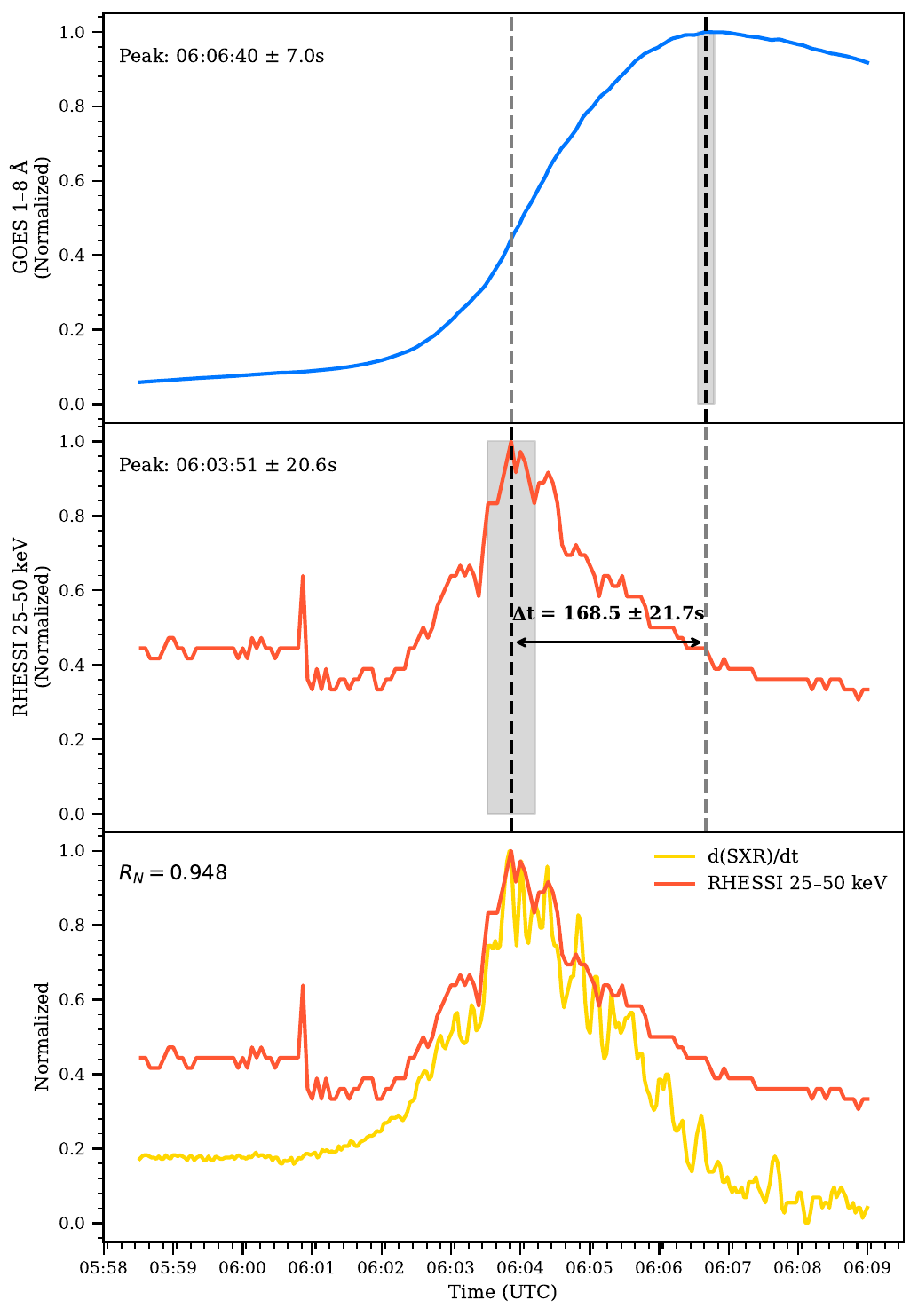}
    \caption{Temporal profiles of a \text{SOL2013-10-26T05:59} (M2.3) flare showing thermal and nonthermal emission.The top panel displays the SXR flux from GOES 1-8 Å. The middle panel shows the RHESSI 25-50 keV HXR light curve. The uncertainty near the peak time is shown in gray shade. The bottom panel compares the time derivative of the GOES 1-8 Å flux with the RHESSI 25-50 keV emission. The cross correlation ($R_{\mathrm{N}}$) for flare M2.3 is 0.948. Delays and cross correlations of the 96 flares are presented in Table~\ref{tab:sample_flares_15}.}
    \label{fig:curves}
\end{figure}

For each event, we selected a time window that covered the entire flare duration. A pre-flare background level was determined using the first few minutes before the event and subtracted from both GOES and RHESSI light curves to isolate the flare emission. The peak time of the GOES 1-8 Å curve was taken as the thermal (SXR) peak, and the peak time of the RHESSI 25-50 keV curve was taken as the nonthermal (HXR) peak. All datasets were aligned in universal time (UT).

To estimate the peak time uncertainties, we use a Monte Carlo approach based only on the measured noise properties of each instrument. The GOES uncertainty is taken as the standard deviation of the pre-flare baseline, while for RHESSI we assume Poisson statistics for the 4-s count rates ($\sigma \approx \sqrt{N}$). These uncertainties quantify the noise in the flux measurements, but they do not directly provide the uncertainty in the peak time, because small fluctuations can shift the location of the maximum in a nonlinear manner. To propagate the real instrumental noise into the timing measurement, we generated 1000 realizations of each light curve by adding random fluctuations consistent with these physically determined uncertainties and recomputed the peak time for each realization. The standard deviation of the resulting distribution was adopted as the 1$\sigma$ uncertainty in each peak time. 
The delay between the two peaks was then calculated as
\begin{equation}
    \Delta t = t_{\mathrm{SXR}} - t_{\mathrm{HXR}},
\end{equation}
This approach enabled us to quantify the time delay, $\Delta $t, between the peak of nonthermal emissions and the thermal emission peak. The total uncertainty in $\Delta $t was obtained by combining the individual peak time errors in quadrature. For a M2.3 flare, as shown in Figure~\ref{fig:curves}, this delay is 168.5 ± 21.7s.

To evaluate the correspondence between the GOES SXR evolution and the RHESSI HXR emission, we computed the time derivative of the GOES 1-8~\AA\ light curve and calculated its Pearson correlation with the RHESSI 25-50~keV light curve. \added{Here we define $R_{\mathrm{N}}$ as this Pearson correlation coefficient, which quantifies how closely the flare
follows the Neupert effect.} The resulting coefficient ($R_{\mathrm{N}}$) provides a purely quantitative measure of the similarity between the two curves. \added{These values were obtained for all selected events; Table~\ref{tab:sample_flares_15} lists the first 15 flares, while the complete results are provided as a machine-readable table online.}

\subsection{Loop Length Determination}
To characterize the flare loop geometry, we used co-temporal imaging from RHESSI and the AIA onboard SDO. The RHESSI imaging time was specifically chosen at the nonthermal peak because this is when the 25-50 keV emission is strongest, ensuring the best signal to noise ratio and most accurate localization of the footpoint sources. For each

\begin{figure}[ht!]
    \centering
    \includegraphics[width=0.65\linewidth]{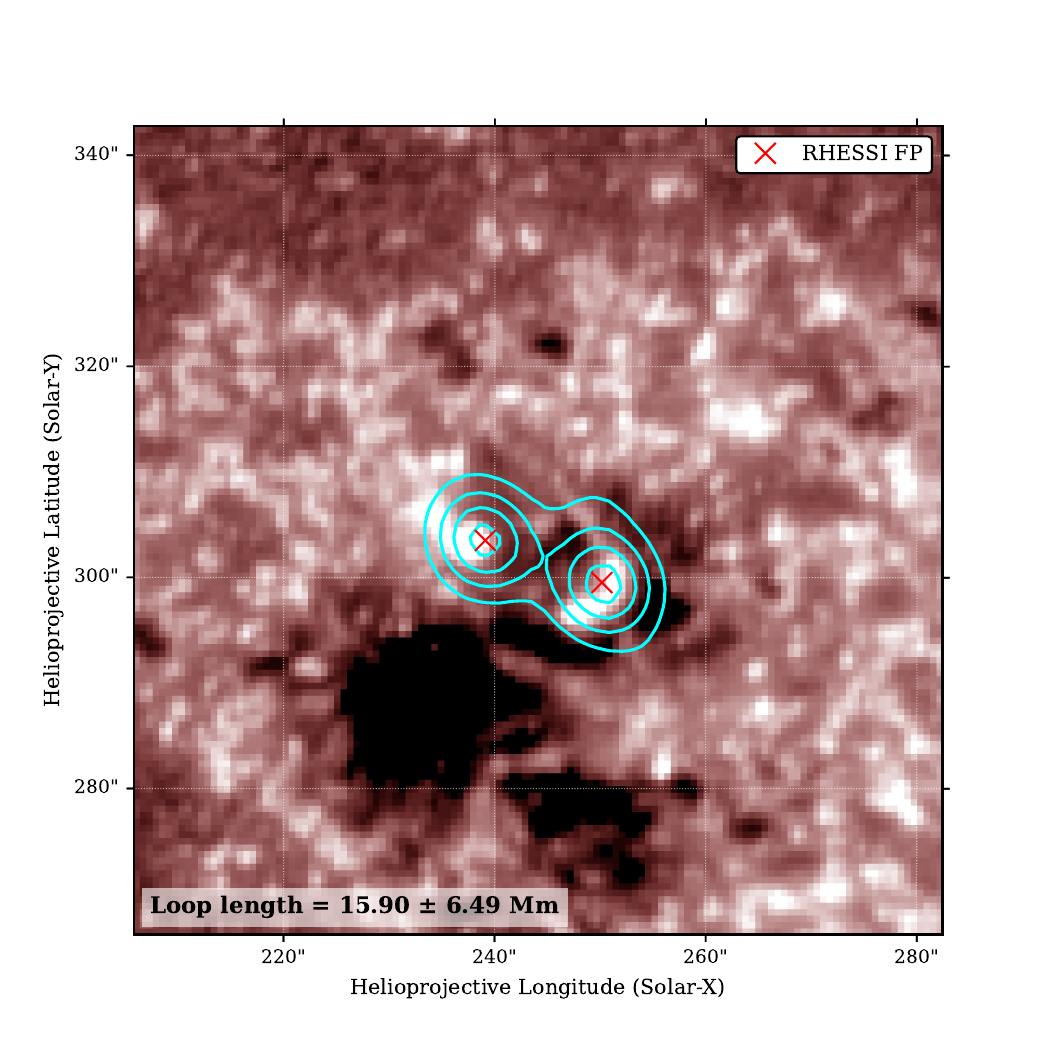}
    \caption{RHESSI 25-50~keV HXR emission contours overlaid on the SDO/AIA 1700~\AA{} image for a \text{SOL2013-04-21T16:00} (C2.9) flare. The two nonthermal footpoints are highlighted in red. The AIA 1700~\AA{} background traces the lower chromospheric emission at the flare site.}
    \label{fig:footpoints}
\end{figure}
 event, we reconstructed a RHESSI CLEAN image in the 25-50 keV band at the time of the nonthermal peak seen in the RHESSI light curve. This energy range corresponds to HXR emission produced by accelerated electrons interacting with the dense chromosphere.

A co-temporal AIA 1700\,\AA\ image was selected to identify the corresponding chromospheric flare ribbons or footpoints. 
The RHESSI and AIA images were co-aligned by matching their solar disk centers and field of view, ensuring that both instruments referred to the same spatial region on the Sun. The two brightest RHESSI 25-50 keV sources were identified as the nonthermal footpoints. These positions were verified against the bright UV emission in the AIA 1700\,\AA\ image to confirm their chromospheric association. The projected footpoint separation distance $D$ was then measured directly from the co-aligned images and converted to physical units using SunPy procedures. Assuming a semicircular loop geometry, the total loop length $L$ was estimated as
\begin{equation}
    L = \frac{\pi D}{2}.
\end{equation}

We note that the semicircular geometry is an approximation; real flare loops may deviate from this shape, leading to either an overestimate or underestimate of the true loop length.
To estimate the uncertainty in the loop length, we adopt the true pointing and resolution limits of the instruments. RHESSI’s best theoretical spatial resolution is 2.26$^{\prime\prime}$ \citep{Hurford2002}, but practical centroiding accuracy is typically 3--4$^{\prime\prime}$, as shown by \citet{DennisPernak2009}. We therefore adopted 4$^{\prime\prime}$ as the positional uncertainty of each RHESSI footpoint. The AIA 1700\,\AA\ images have a pixel scale of 0.6$^{\prime\prime}$ and a diffraction–limited resolution of about 1.5$^{\prime\prime}$ \citep{Lemen2012AIA}. After reprojecting the RHESSI CLEAN images onto the AIA grid, the combined positional uncertainty of the footpoints is obtained by adding the RHESSI (4$^{\prime\prime}$) and AIA (1.5$^{\prime\prime}$) uncertainties in quadrature, giving approximately 4.3$^{\prime\prime}$. Converting this to AIA pixels (4.3$^{\prime\prime}$/0.6$^{\prime\prime}$) yields $\sim$7 pixels, which we adopt as the physically justified perturbation scale in the Monte Carlo loop-length calculation. Each footpoint is therefore perturbed within this $\pm7$\,pixel uncertainty over 1,000 trials, the loop length is recomputed, and the mean and standard deviation of the resulting distribution provide the final loop length and its 1$\sigma$ uncertainty. Figure~\ref{fig:footpoints} shows the RHESSI contours overlaid on the AIA 1700\,\AA\ image, and using Equation~(2), we obtained a loop length of $15.90 \pm 6.49$~Mm for this C2.9 flare. Similar measurements were performed for all selected events, and the results are summarized in Table~\ref{tab:sample_flares_15}.

\begin{table*}[ht]
\centering
\caption{First 15 flares from the analyzed dataset. The full catalog is available as machine-readable table online.}
\small
\setlength{\tabcolsep}{4pt}
\begin{tabular}{cccccccccc}
\hline\hline
SOL ID & Class & X($''$) & Y($''$) & HXR Peak & SXR Peak & $\Delta t$ (s) & Loop length (Mm) & $R_{\mathrm{N}}$ \\
\hline
SOL2014-01-04T10:16 & M1.3 & $-704.4$ & $-78.0$   & 10:20:24$\pm$21.53 & 10:25:35$\pm$27.06 & 311$\pm$34.58 & 47.44$\pm$9.61 & 0.752 \\
SOL2014-01-07T10:07 & M7.2 & $-83.9$  & $-40.9$   & 10:11:20$\pm$12.89 & 10:13:19$\pm$5.30  & 119$\pm$13.94 & 28.46$\pm$6.91 & 0.819 \\
SOL2014-02-02T06:24 & M2.6 & $-182.8$ & $290.2$   & 06:33:56$\pm$3.98  & 06:34:15$\pm$22.27 & 19$\pm$22.62   & 14.67$\pm$5.09 & 0.261 \\
SOL2014-02-02T18:05 & M3.1 & $-66.1$  & $-93.7$   & 18:10:36$\pm$22.91 & 18:11:49$\pm$3.93  & 73$\pm$23.24   & 14.27$\pm$4.99 & 0.867 \\
SOL2014-02-06T22:56 & M1.5 & $725.8$  & $-137.3$  & 23:02:46$\pm$3.40  & 23:04:59$\pm$8.00  & 133$\pm$8.69   & 17.01$\pm$5.32 & 0.562 \\
SOL2014-02-07T10:25 & M1.9 & $767.0$  & $226.2$   & 10:28:12$\pm$2.89  & 10:29:21$\pm$2.00  & 69$\pm$3.51    & 19.60$\pm$6.81 & 0.838 \\
SOL2014-02-12T03:52 & M3.7 & $16.6$   & $-77.2$   & 04:21:00$\pm$12.60 & 04:25:27$\pm$10.0  & 267$\pm$16.09  & 38.64$\pm$4.81 & 0.743 \\
SOL2014-02-12T15:41 & M2.1 & $115.8$  & $-78.0$   & 15:49:28$\pm$9.42  & 15:50:57$\pm$6.04  & 89$\pm$11.19   & 23.36$\pm$11.62 & 0.834 \\
SOL2014-02-13T15:45 & M1.4 & $329.7$  & $-0.2$    & 15:55:16$\pm$27.6  & 15:57:09$\pm$9.00  & 113$\pm$29.03  & 24.30$\pm$4.99 & 0.582 \\
SOL2014-02-14T12:29 & M1.6 & $474.9$  & $-94.1$   & 12:34:36$\pm$2.40  & 12:40:23$\pm$4.00  & 347$\pm$4.66   & 54.12$\pm$5.13 & 0.794 \\
SOL2014-03-29T17:35 & X1.0 & $472.9$  & $229.7$   & 17:45:40$\pm$30.32 & 17:48:25$\pm$11.59 & 165$\pm$32.46  & 24.49$\pm$5.69 & 0.865 \\
SOL2014-04-18T12:31 & M7.3 & $513.3$  & $-233.6$  & 12:54:04$\pm$57.91 & 13:03:38$\pm$28.80 & 574$\pm$64.68  & 41.09$\pm$5.37 & 0.874 \\
SOL2014-06-12T04:14 & M2.0 & $-717.3$ & $-333.4$  & 04:19:24$\pm$11.93 & 04:21:27$\pm$39.16 & 123$\pm$40.94  & 25.38$\pm$9.89 & 0.883 \\
SOL2014-09-23T23:03 & M2.3 & $-551.4$ & $-248.3$  & 23:09:16$\pm$14.18 & 23:16:45$\pm$10.91 & 449$\pm$17.89  & 44.09$\pm$4.95 & 0.804 \\
SOL2014-09-28T02:39 & M5.1 & $415.4$  & $-343.4$  & 02:46:28$\pm$24.86 & 02:58:42$\pm$28.05 & 734$\pm$37.48  & 65.34$\pm$12.18 & 0.759 \\
\hline
\end{tabular}
\label{tab:sample_flares_15}
\end{table*}

\section{Results}

\subsection{ Flare Loop Length vs Emission Peak Delay}
\begin{figure}[ht!]
    \centering
    \includegraphics[width=0.65\linewidth]{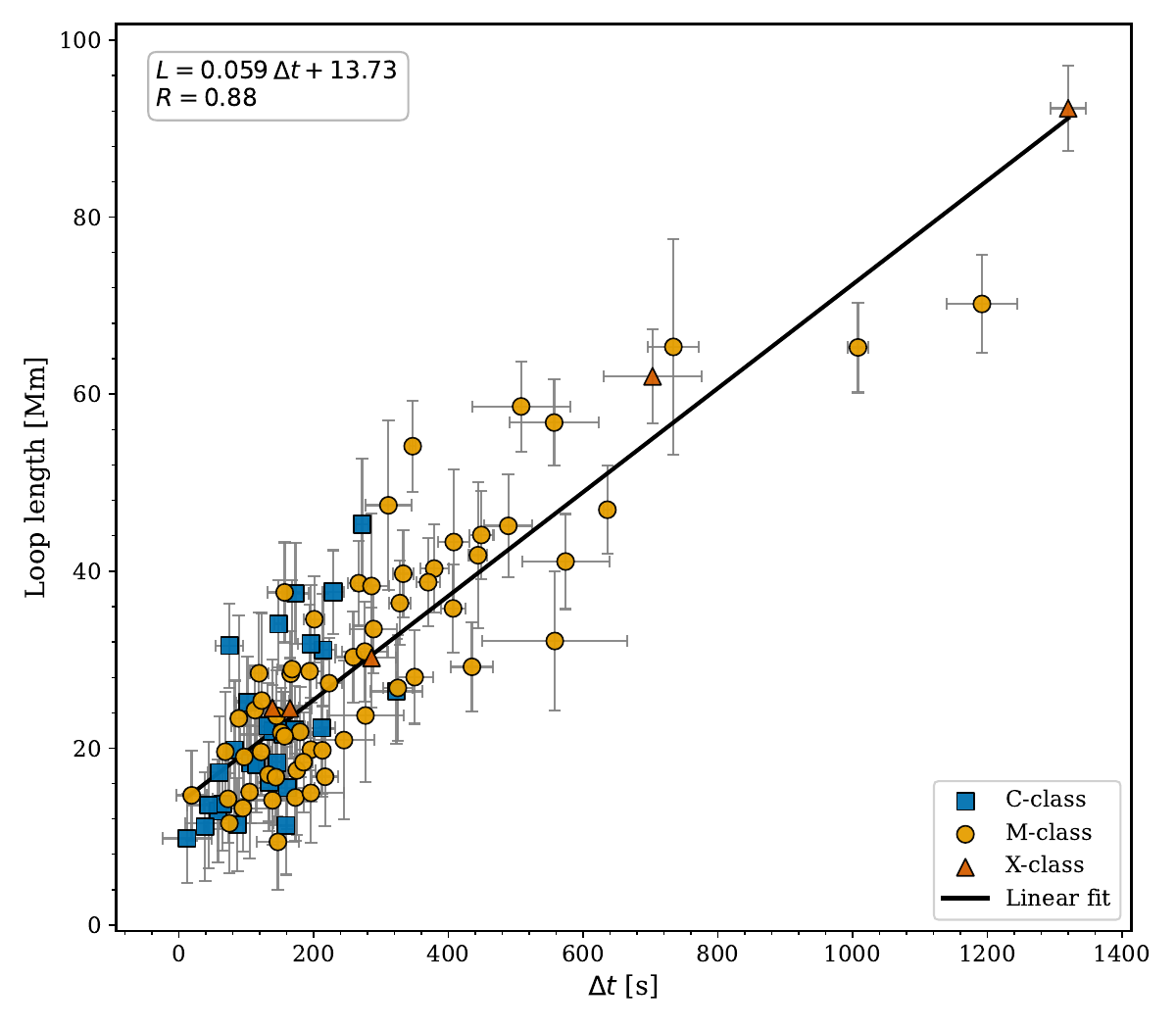}
    \caption{Magnetic loop length versus the HXR–SXR time delay $\Delta t$ for all 96 flares, color-coded by GOES class.}
    \label{fig:all_flares}
\end{figure}

Figure~\ref{fig:all_flares} shows the relationship between the magnetic loop length ($L$) and the temporal delay $\Delta t$ for all 96 solar flares in our dataset. The sample includes flares spanning the full GOES classification range, from C-class (square) to M-class (circle) and X-class (triangle). Each data point represents a single flare, with horizontal and vertical error bars indicating the uncertainties in $\Delta t$ and $L$, respectively.

A clear positive trend is evident: flares with longer magnetic loops consistently exhibit larger values of $\Delta t$. A linear fit to the complete unfiltered sample yields

\begin{equation}
    L \;=\; 0.059\,\Delta t \;+\; 13.73\quad\text{(Mm)},
\end{equation}

with a Pearson correlation coefficient of $R = 0.88$. This result indicates that, even without applying any physical filtering or selection criteria, the dataset already reveals a well defined global relationship between the loop length and the time delay between the SXR and HXR emission peaks.

\subsection{Subset of Flares Selected by $R_{\mathrm{N}}$ Threshold}

To examine how the trend behaves for events with different levels of correspondence between thermal and nonthermal diagnostics, we defined subsets of flares using correlation thresholds between the GOES 1--8~\AA\ derivative and the RHESSI 25--50~keV light curve. We first selected all flares with a correlation ($R_{\mathrm{N}}$) of at least 0.5, yielding a subset of 87 events. A linear fit to this selected sample gives

\begin{figure}[ht!]
    \centering
    \includegraphics[width=0.65\linewidth]{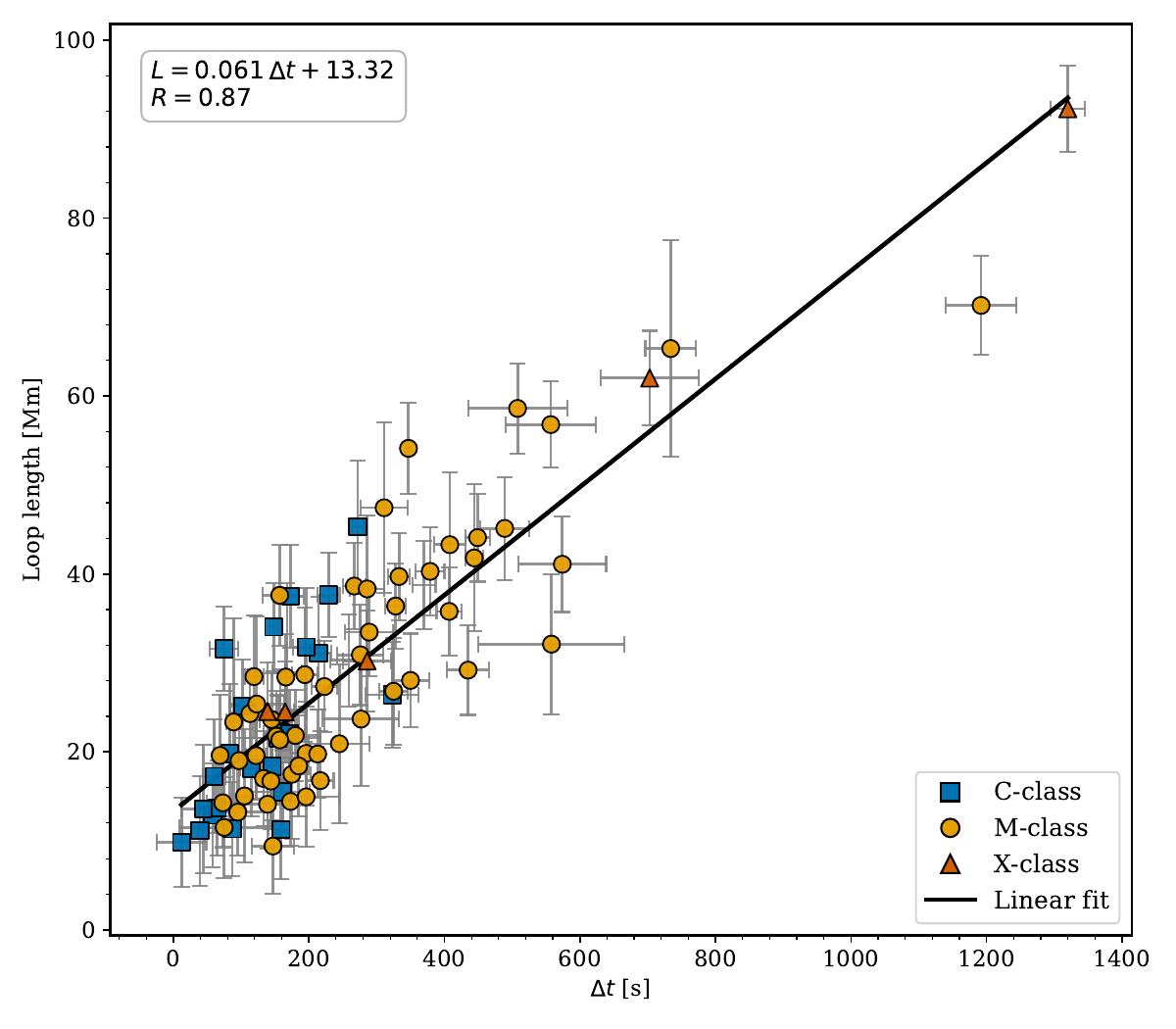}
    \caption{Same as Figure 4, but only for flares with \added{Neupert} correlation ($R_{\mathrm{N}}$) $\geq 0.5$ (87 events). }
    \label{fig:87}
\end{figure}

\begin{equation}
    L \;=\; 0.061\,\Delta t \;+\; 13.32 \quad \text{(Mm)},
\end{equation}
with a correlation of $R = 0.87$ as shown in figure~\ref{fig:87}. This trend is very similar to that of the full unfiltered sample, indicating that the global $L$--$\Delta t$ relationship remains stable when the analysis is restricted to  flares with $R_{\mathrm{N}} \geq 0.5$.

\begin{figure}[ht!]
    \centering
    \includegraphics[width=0.65\linewidth]{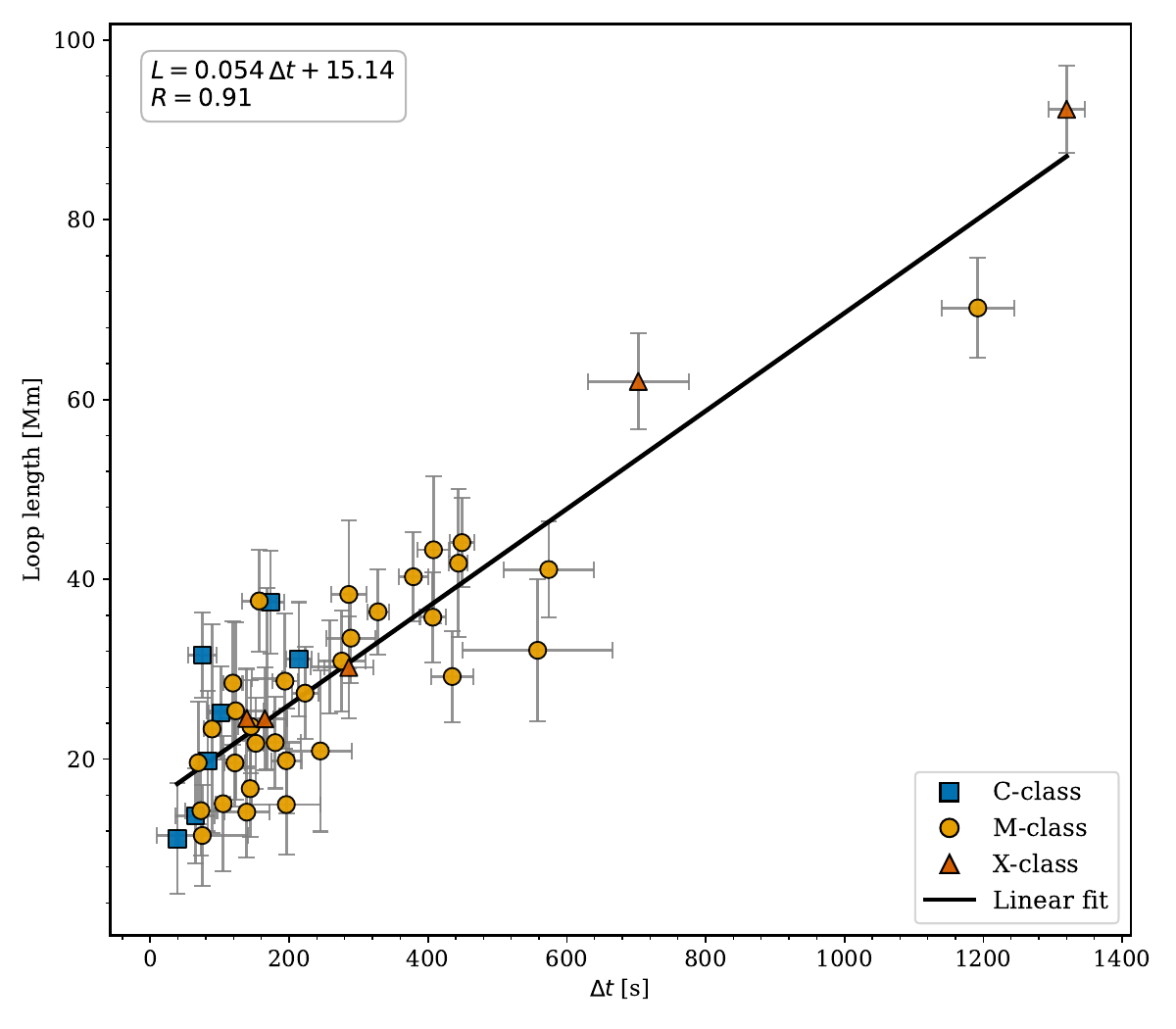}
    \caption{Same as Figure 4, but limited to flares with \added{Neupert} correlation ($R_{\mathrm{N}}$) $\geq 0.8$ (46 events). }
    \label{fig:46}
\end{figure}

We then applied a stricter condition of correlation ($R_{\mathrm{N}}$) $\geq 0.8$, resulting in 46 high confidence flares. For this refined subset, the relationship between loop length and delay becomes even tighter. A linear fit gives

\begin{equation}
    L \;=\; 0.054\,\Delta t \;+\; 15.14 \quad \text{(Mm)},
\end{equation}

with an enhanced correlation of $R = 0.91$ as shown in figure~\ref{fig:46}. 

\section{Discussion}

The linear fits presented in Section~4 (Equations~3--5) reveal a strong and consistent relationship between the magnetic loop length ($L$) and the time delay ($\Delta t$) between the HXR and SXR peaks, with slopes ranging from \added{$54$–$61~\mathrm{km~s^{-1}}$.} As we progressively select flares with stronger Neupert behaviour, the correlation coefficient increases from $R = 0.88$ for the full sample to $R = 0.91$ for the most Neupert consistent events. We characterize the ``\added{Neupertian}''\footnote{The term \emph{Neupertian} was coined by Hugh Hudson and first appeared in \url{https://heliowiki.smce.nasa.gov/wiki/index.php/Neupertianity}.}
behaviour of a flare by the Pearson coefficient ($R_{\mathrm{N}}$); the closer it is to 1, the more \added{Neupertian} the flare is \citep{DennisZarro1993}.
 Moreover, we adopt the canonical interpretation of the Neupert effect \citep{Neupert1968}, i.e., the nonthermal particles transfer their kinetic energy to the chromospheric plasma, which ``evaporates'' and fills the magnetic loops emitting SXR. With these hypotheses in mind we can conclude that the timing of flare emission is closely linked to the underlying magnetic geometry, and that this link becomes clearer when energy transport is dominated by impulsive, beam-driven heating. In the strongly Neupertian subset (($R_{\mathrm{N}}$) $\geq 0.8$), the delay between HXR and SXR peaks can be directly interpreted as the loop filling time required for evaporated plasma to reach its maximum thermal emission. Conversely, flares with weaker Neupert correlations ($R_{\mathrm{N}} < 0.5$) likely involve additional heating processes such as conductive pre-heating, multi-threaded or extended reconnection, and gradual energy release which distort the simple HXR--SXR timing relationship and reduce the strength of the correlation \citep[e.g.][]{Veronig2002b, WarmuthMann2016}. The stability of the slope across all subsets, combined with the improvement in correlation for more Neupertian events, reinforces that the observed $L$-$\Delta t$ relationship primarily reflects the physics of chromospheric evaporation driven by nonthermal electrons \citep{DennisZarro1993, WarmuthMann2016}.

\added{The slopes in our linear fits (54--61~km~s$^{-1}$) describe the overall
trend between loop length and delay time. Chromospheric expansion is a hydrodynamic response, whereas the observed X-ray emission results from Coulomb interactions, and the timing of these processes need not be strictly identical. In this sense, the fitted slopes provide a characteristic, average scale proxy for evaporation related filling rates across the flare sample. However, because the linear fits include non-zero intercepts, the average filling speed for a given event cannot be taken directly from the slope alone.  When taking loop lengths predicted by the fits for each time delay shows that flares with short delays (less than about 300~s) typically show average filling speeds of about 130-250~km~s$^{-1}$, with a few compact events reaching values above 300~km~s$^{-1}$. For long duration events (greater than about 300~s), the speeds are much smaller, typically around 70-100~km~s$^{-1}$. These values represent averages over the full SXR rise phase, so the true evaporation flows during the early impulsive phase are likely higher.} \added{The average speed value obtained from the fits} is consistent with the gentle to explosive chromospheric evaporation regime described by \citet{Fisher1985a, Fisher1985b}, who showed that for energy fluxes below about $10^{10}~\mathrm{erg~cm^{-2}~s^{-1}}$, upflows remain under $30~\mathrm{km~s^{-1}}$, while higher fluxes generate explosive evaporation with velocities reaching several hundred $\mathrm{km~s^{-1}}$. Our derived speeds therefore suggest that most of the analyzed events lie near the transition between these two regimes. Observational results by \citet{MilliganDennis2009} and \citet{Sadykov2019} showed \ion{Fe}{21}-\ion{Fe}{24} blueshifts of $200$-$250~\mathrm{km~s^{-1}}$ for hot ($10$--$16~\mathrm{MK}$) upflows and \ion{Si}{4} redshifts of $30$-$40~\mathrm{km~s^{-1}}$ for cooler condensation layers \citep[e.g.][]{Graham2015,Polito2016,Graham2020}, consistent with our inferred velocities when considering projection and temporal averaging. Theoretical simulations by \citet{Reep2015} also demonstrated that electron beams with fluxes of $10^{9}$-$10^{11}~\mathrm{erg~cm^{-2}~s^{-1}}$ can drive upflows ranging from $100$ to over $1000~\mathrm{km~s^{-1}}$ depending on beam energy, reinforcing that the observed $L$-$\Delta t$ scaling likely reflects the hydrodynamic response time of plasma expansion along the loop. In this framework, longer loops require a longer time for evaporated material to reach equilibrium, producing larger delays between the HXR and SXR peaks.

Further support for this interpretation comes from \citet{Polito2018}, whose RADYN simulations of both electron beam and conduction driven heating produced similar upflow velocities of $200$-$500~\mathrm{km~s^{-1}}$ and downflows of $50~\mathrm{km~s^{-1}}$, demonstrating that chromospheric evaporation is an inevitable response to impulsive heating. Events with weaker Neupert behavior likely include a combination of continued reconnection, extended coronal heating, or delayed thermal conduction, which can prolong the SXR emission independently of the chromospheric filling timescale \citep{ReepToriumi2017}.
The intercept of 13-15~Mm should not be interpreted as a physical lower limit to the loop length. It most likely reflects systematic effects such as early heating and limited timing accuracy. In some flares, energy released in the corona can be transported by thermal conduction or low energy electrons before strong HXR emission develops, producing an early SXR rise not perfectly tracked by the HXR curve \citep[e.g.][]{Reep2015, Polito2018}. This pre-peak thermal response, together with uncertainties in defining the exact SXR and HXR maxima especially in complex or multi-threaded event can yield an apparent positive offset in the $L$-$\Delta t$ relation. Consistent with this interpretation, previous Neupert effect studies reported minute scale offsets between the HXR onset, the SXR derivative, and the SXR maximum \citep[e.g.][]{Veronig2002b, WarmuthMann2016}.

Although the $L$-$\Delta t$ trend is statistically strong, a few limitations should be acknowledged. First, our analysis assumes a single representative loop per flare and models it with a semicircular geometry, even though real flares may involve many smaller loops that are heated in sequence. Second, the determination of peak times from GOES and RHESSI light curves is affected by background subtraction and instrument cadence, which can introduce additional uncertainty, especially for weaker C-class events. Third, we use only one HXR energy band (25-50~keV) to represent nonthermal electrons, even though the most suitable energy range can vary depending on the flare. These factors may contribute to some of the scatter in the timing measurements but do not change the overall trend we observe.

\section{Conclusions}

We have conducted a statistical, imaging-based study of 96 solar flares observed between 2013 and 2015 using GOES, RHESSI, and SDO/AIA. We find that the delay between HXR and SXR peaks scales linearly with the magnetic loop length, with slopes in the range \added{$54$–$61~\mathrm{km~s^{-1}}$} and intercepts of $13$-$15~\mathrm{Mm}$. The correlation coefficient increases from $R = 0.88$ for the full sample to $R = 0.91$ for strongly Neupert like events, showing that the relationship is most clearly expressed when nonthermal electrons dominate the energy input. \added{The inferred average evaporation velocities for individual flares range from $70$-$250~\mathrm{km\,s^{-1}}$, placing the events within the gentle to explosive chromospheric evaporation regime and indicating that the observed delay traces the efficiency of energy transport and loop filling.}
 Our results provide direct, imaging based evidence that the temporal evolution of solar flares is controlled by their magnetic geometry.

Future work will test this interpretation on an event by event basis by relaxing the semicircular loop assumption, incorporating multi-threaded structures, and comparing with hydrodynamic and radiative simulations tailored to reproduce the observed $L$-$\Delta t$ scaling. \added{In addition, flares with small $R_N$ values will be studied more closely, because events that do not show a clear Neupertian pattern may involve extra heating processes or more complicated energy transport.}

\section{Software and third party data repository citations}

Portions of the text were edited for clarity using ChatGPT (OpenAI, 2025)(\url{https://openai.com/chatgpt}). 

\section*{Acknowledgements}

This work was partially supported by FAPESP (grant 2022/15700-7) and CAPES-PRINT (grant 88887.310385/2018-00). 
C.~G.~G.~C. is a corresponding researcher of CONICET (Argentina) and a CNPq research fellow (PQ level 1, grant 302836/2022-5). 
P.~J.~A.~S. acknowledges support from CNPq (grant 305808/2022-2) and MackPesquisa (grant 231017). 
S.~M.~P. acknowledges full financial support from the Coordenação de Aperfeiçoamento de Pessoal de Nível Superior – Programa de Suporte à Pós-Graduação de Instituições Comunitárias de Educação Superior (CAPES PROSEC, modality I–2024). 
S.~S.~S. acknowledges support from the Conselho Nacional de Desenvolvimento Científico e Tecnológico (CNPq), Brazil, through a long-duration Master's fellowship (Processo 134620/2025-9). We thank the anonymous referee for the constructive comments and suggestions, which helped improve the clarity and quality of this manuscript.

\bibliography{sample701}{}

@article{PriestForbes2002,
  author  = {Priest, E.~R. and Forbes, T.~G.},
  title   = {The Magnetic Nature of Solar Flares},
  journal = {The Astronomy and Astrophysics Review},
  year    = {2002},
  volume  = {10},
  number  = {4},
  pages   = {313--377},
  doi     = {10.1007/s001590100013}
}

@article{LinForbes2000,
  author  = {Lin, J. and Forbes, T. G.},
  title   = {Effects of reconnection on the coronal mass ejection process},
  journal = {Journal of Geophysical Research: Space Physics},
  year    = {2000},
  volume  = {105},
  number  = {A2},
  pages   = {2375--2392},
  doi     = {10.1029/1999JA900477}
}

@article{Brown1971,
  author  = {Brown, J. C.},
  title   = {The Deduction of Energy Spectra of Non-Thermal Electrons in Flares from the Observed Dynamic Spectra of Hard X-Ray Bursts},
  journal = {Solar Physics},
  year    = {1971},
  volume  = {18},
  pages   = {489--502},
  doi     = {10.1007/BF00149070}
}

@article{Emslie1978,
  author  = {Emslie, A.~G.},
  title   = {The Collisional Interaction of a Beam of Charged Particles with a Hydrogen Target of Arbitrary Ionization Level},
  journal = {The Astrophysical Journal},
  year    = {1978},
  volume  = {224},
  pages   = {241--246},
  doi     = {10.1086/156371}
}

@article{DennisZarro1993,
  author  = {Dennis, B.~R. and Zarro, D.~M.},
  title   = {The Neupert Effect: What Can It Tell Us About the Impulsive and Gradual Phases of Solar Flares},
  journal = {Solar Physics},
  year    = {1993},
  volume  = {146},
  number  = {1},
  pages   = {177--190},
  doi     = {10.1007/BF00662178}
}

@article{Neupert1968,
  author  = {Neupert, W.~M.},
  title   = {Comparison of Solar X-Ray Line Emission with Microwave Emission during Flares},
  journal = {The Astrophysical Journal},
  year    = {1968},
  volume  = {153},
  pages   = {L59--L64},
  doi     = {10.1086/180220}
}

@article{Veronig2002a,
  author  = {Veronig, A.~M. and Temmer, M. and Hanslmeier, A. and Otruba, W. and Messerotti, M.},
  title   = {Temporal aspects and frequency distributions of solar soft X-ray flares},
  journal = {A\&A},
  volume  = {382},
  pages   = {1070--1080},
  year    = {2002},
  doi     = {10.1051/0004-6361:20011694}
}

@article{LiGan2006,
  author  = {Li, Y. P. and Gan, W. Q.},
  title   = {On the Peak Times of Thermal and Nonthermal Emissions in Solar Flares},
  journal = {Astrophysical Journal Letters},
  year    = {2006},
  volume  = {652},
  pages   = {L61--L64},
  doi     = {10.1086/509879}
}

@article{Aschwanden1998,
  author  = {Aschwanden, M.~J. and Kliem, B. and Schwarz, U. and Kurths, J. and Dennis, B.~R. and Schwartz, R.~A.},
  title   = {Wavelet Analysis of Solar Flare Hard X-Rays},
  journal = {The Astrophysical Journal},
  year    = {1998},
  volume  = {505},
  number  = {2},
  pages   = {941--956},
  doi     = {10.1086/306200}
}

@article{Nagai1980,
  author  = {Nagai, F.},
  title   = {A Model of Hot Loops Associated with Solar Flares. Part One: Gasdynamics in the Loops},
  journal = {Solar Physics},
  year    = {1980},
  volume  = {68},
  number  = {2},
  pages   = {351--379},
  doi     = {10.1007/BF00156874}
}

@article{Fisher1985a,
  author  = {Fisher, G.~H. and Canfield, R.~C. and McClymont, A.~N.},
  title   = {Flare Loop Radiative Hydrodynamics. VII. Dynamics of the Thick Target Heated Chromosphere},
  journal = {The Astrophysical Journal},
  year    = {1985},
  volume  = {289},
  pages   = {434--455},
  doi     = {10.1086/162903}
}

@article{Fisher1985b,
  author  = {Fisher, G.~H. and Canfield, R.~C. and McClymont, A.~N.},
  title   = {Flare Loop Radiative Hydrodynamics. V. Response to Thick-Target Heating},
  journal = {The Astrophysical Journal},
  year    = {1985},
  volume  = {289},
  pages   = {414--424},
  doi     = {10.1086/162901}
}

@article{Mariska1989,
  author  = {Mariska, J.~T. and Emslie, A.~G. and Li, P.},
  title   = {Numerical Simulations of Impulsively Heated Solar Flares},
  journal = {The Astrophysical Journal},
  year    = {1989},
  volume  = {341},
  pages   = {1067--1074},
  doi     = {10.1086/167564}
}

@article{Pesnell2012,
  author  = {Pesnell, W.~D. and Thompson, B.~J. and Chamberlin, P.~C.},
  title   = {The Solar Dynamics Observatory (SDO)},
  journal = {Solar Physics},
  year    = {2012},
  volume  = {275},
  number  = {1-2},
  pages   = {3--15},
  doi     = {10.1007/s11207-011-9841-3}
}

@article{Lin2002RHESSI,
  author  = {Lin, R.~P. and Dennis, B.~R. and Hurford, G.~J. and Smith, D.~M. and Zehnder, A. and Harvey, P.~R. and Curtis, D.~W. and Pankow, D. and Turin, P. and Bester, M. and Csillaghy, A. and {et al.}},
  title   = {The Reuven Ramaty High-Energy Solar Spectroscopic Imager (RHESSI)},
  journal = {Solar Physics},
  year    = {2002},
  volume  = {210},
  number  = {1-2},
  pages   = {3--32},
  doi     = {10.1023/A:1022428818870}
}

@article{Lemen2012AIA,
  author  = {Lemen, J.~R. and Title, A.~M. and Akin, D.~J. and Boerner, P.~F. and Chou, C. and Drake, J.~F. and Duncan, D.~W. and Edwards, C.~G. and Friedlaender, F.~M. and Heyman, G.~F. and Hurlburt, N.~E. and Katz, N.~L. and Kushner, G.~D. and Levay, M. and Lindgren, R.~W. and Mathur, D.~P. and McFeaters, E.~L. and Mitchell, S. and Rehse, R.~A. and Schrijver, C.~J. and {et al.}},
  title   = {The Atmospheric Imaging Assembly (AIA) on the Solar Dynamics Observatory (SDO)},
  journal = {Solar Physics},
  year    = {2012},
  volume  = {275},
  number  = {1-2},
  pages   = {17--40},
  doi     = {10.1007/s11207-011-9776-8}
}

@article{CargillMariskaAntiochos1995,
  author  = {Cargill, P.~J. and Mariska, J.~T. and Antiochos, S.~K.},
  title   = {Cooling of Solar Flare Plasmas. I. Theoretical Considerations},
  journal = {The Astrophysical Journal},
  year    = {1995},
  volume  = {439},
  pages   = {1034--1043},
  doi     = {10.1086/175240}
}

@article{Veronig2002b,
  author  = {Veronig, A.~M. and Vršnak, B. and Temmer, M. and Hanslmeier, A. and Messerotti, M.},
  title   = {Temporal relationship between soft and hard X-ray emissions of solar flares},
  journal = {A\&A},
  volume  = {392},
  pages   = {699--712},
  year    = {2002},
  doi     = {10.1051/0004-6361:20020947}
}

@article{WarmuthMann2016,
  author = {Warmuth, A. and Mann, G.},
  title = {Constraints on Energy Release in Solar Flares from Multiwavelength Observations},
  journal = {A\&A},
  year = {2016},
  volume = {588},
  pages = {A116},
  doi = {10.1051/0004-6361/201527474}
}

@article{MilliganDennis2009,
  author = {Milligan, R. O. and Dennis, B. R.},
  title = {Velocity Characteristics of Evaporation in Solar Flares},
  journal = {ApJ},
  year = {2009},
  volume = {699},
  pages = {968--975},
  doi = {10.1088/0004-637X/699/1/968}
}

@article{Polito2018,
  author = {Polito, V. and Testa, P. and Allred, J. and De Pontieu, B. and Carlsson, M. and Pereira, T. M. D. and Gosic, M. and Reale, F.},
  title = {Investigating the Response of Loop Plasma to Nanoflare Heating Using RADYN Simulations},
  journal = {The Astrophysical Journal},
  year = {2018},
  volume = {856},
  number = {2},
  pages = {178},
  doi = {10.3847/1538-4357/aab49e}
}

@article{Sadykov2019,
  author  = {Sadykov, V. M. and Kosovichev, A. G. and Sharykin, I. N. and Kerr, G. S.},
  title   = {Statistical Study of Chromospheric Evaporation in the Impulsive Phase of Solar Flares},
  journal = {The Astrophysical Journal},
  year    = {2019},
  volume  = {871},
  pages   = {2},
  doi     = {10.3847/1538-4357/aaf6b0}
}

@article{Reep2015,
  author  = {Reep, J. W. and Bradshaw, S. J. and Alexander, D.},
  title   = {Optimal Electron Energies for Driving Chromospheric Evaporation in Solar Flares},
  journal = {The Astrophysical Journal},
  year    = {2015},
  volume  = {808},
  pages   = {177},
  doi     = {10.1088/0004-637X/808/2/177}
}

@article{ReepToriumi2017,
  author  = {Reep, J. W. and Toriumi, S.},
  title   = {The Direct Relation between the Duration of Magnetic Reconnection and the Evolution of GOES Light Curves in Solar Flares},
  journal = {The Astrophysical Journal},
  year    = {2017},
  volume  = {851},
  pages   = {4},
  doi     = {10.3847/1538-4357/aa96fe}
}

@article{Graham2015,
  author  = {Graham, D.~R. and Cauzzi, G.},
  title   = {Temporal Evolution of Multiple Evaporating Ribbon Sources in a Solar Flare},
  journal = {ApJL},
  volume  = {807},
  number  = {2},
  pages   = {L22},
  year    = {2015},
  doi     = {10.1088/2041-8205/807/2/L22}
}

@article{Polito2016,
  author  = {Polito, V. and Reep, J.~W. and Reeves, K.~K. and Sim{\~o}es, P.~J.~A. and Dud{\'i}k, J. and Del Zanna, G. and Mason, H.~E. and Golub, L.},
  title   = {Simultaneous IRIS and Hinode/EIS Observations and Modeling of the 2014 October 27 X2.0 Class Flare},
  journal = {ApJ},
  volume  = {816},
  number  = {2},
  pages   = {89},
  year    = {2016},
  doi     = {10.3847/0004-637X/816/2/89}
}

@article{Graham2020,
  author  = {Graham, D.~R. and Cauzzi, G. and Zangrilli, L. and Kowalski, A. and Sim{\~o}es, P.~J.~A. and Allred, J.},
  title   = {Spectral Signatures of Chromospheric Condensation in a Major Solar Flare},
  journal = {ApJ},
  volume  = {895},
  number  = {1},
  pages   = {6},
  year    = {2020},
  doi     = {10.3847/1538-4357/ab88ad}
}

@article{Hurford2002,
  author  = {Hurford, G. J. and Schmahl, E. J. and Schwartz, R. A. and others},
  title   = {The RHESSI Imaging Concept},
  journal = {Solar Physics},
  volume  = {210},
  number  = {1},
  pages   = {61--86},
  year    = {2002},
  doi     = {10.1023/A:1022436213688},
  adsurl  = {https://ui.adsabs.harvard.edu/abs/2002SoPh..210...61H},
}

@article{DennisPernak2009,
  author  = {Dennis, B. R. and Pernak, R. L.},
  title   = {Hard X-Ray Flare Source Sizes Measured with the Ramaty High Energy Solar Spectroscopic Imager},
  journal = {The Astrophysical Journal},
  volume  = {698},
  number  = {2},
  pages   = {2131--2143},
  year    = {2009},
  doi     = {10.1088/0004-637X/698/2/2131},
  adsurl  = {https://ui.adsabs.harvard.edu/abs/2009ApJ...698.2131D},
}
\bibliographystyle{aasjournalv7}



\end{document}